\documentclass[aps,prb,onecolumn,floatfix]{revtex4-2}

\usepackage{epsf}
\usepackage{subfigure}
\usepackage{amsmath}
\usepackage{amssymb}
\usepackage{bm}
\usepackage{graphicx}
\usepackage{epstopdf}
\newcommand{\be}{\begin{equation}}
\newcommand{\ee}{\end{equation}}
\newcommand{\bea}{\begin{eqnarray}}
\newcommand{\eea}{\end{eqnarray}}

\begin{document}

\title{\bf Revealing hidden structures and symmetries in nonequilibrium transport}

\author{David Andrieux}

\begin{abstract}
Recent results have shown how to partition the space of Markov systems into dynamical equivalence classes. 
These equivalence classes structure transport properties in a way that makes, among other features, their responses fully symmetric.
In this note, I illustrate this approach on two representative systems.
First, I derive analytical expressions for the equivalence classes of a disordered ring model.
Second, I verify on a model of ion transport that, within an equivalence class, the response of coupled currents is symmetric both near and far from equilibrium.
\end{abstract}

\maketitle

\vskip 0,25 cm

\section*{The Challenge: Nonequilibrium Response is ill-defined}

Nonequilibrium transport is traditionally characterized by quantities such as currents, affinities or the entropy production of the system. 
Current fluctuations and related fluctuation theorems also provide additional constraints and insights on the behavior of nonequilibrium systems.

However, despite their relevance, these concepts are {\it not enough to build a well-defined nonequilibrium theory}. What do I mean by that? 

The core issue is that specifying a system's currents and affinities (and thus its entropy production) does not uniquely characterize its nonequilibrium state. Multiple combinations of parameters can lead to the same affinities and currents \cite{FN01}. 
As a result, you can drive a system out of equilibrium in an infinite number of ways starting from a given equilibrium state.
While in the linear response regime this ambiguity doesn't really matter since all response coefficients are identical regardless of the path you choose, this is no longer true in the nonlinear regime \cite{H05}. 

Conversely, imagine now that you start with a system out of equilibrium.
A continuum of equilibrium states could serve as reference point and, without a clearly identified equilibrium state and path to it, the nonequilibrium response of the system cannot be uniquely defined.

So, how do we remove this arbitrariness and connect equilibrium and nonequilibrium dynamics in a way that is: 
\begin{itemize}
    \item {\it Well-defined}, i.e. the system's parameters uniquely map all possible nonequilibrium conditions 
    \item {\it Intrinsic}, i.e. the connection emerges from the dynamical characteristics of the system itself.
\end{itemize}

To structure the space of possible dynamics while satisfying these requirements, I previously introduced the concept of {\it dynamical equivalence classes} \cite{A12b, A12c}. 
These 'hidden structures' uniquely characterize the nonquilibrium conditions and parameters of the system, and describe a well-defined nonequilibrium theory.
In addition, using these structures the current fluctuations can be obtained from an equilibrium dynamics and their nonlinear responses take an especially simple, symmetric form \cite{A22}.
These properties go beyond the traditional current fluctuation theorem \cite{A22, A12b}, offering a new lens to understand nonequilibrium systems.

To build physical intuition around dynamical equivalence classes, 
I recently derived their structure and corresponding transport properties for a molecular motor toy model \cite{A23c}.
In this note, my objective is to provide additional examples of how this framework plays out in two specific cases. 
I first provide an analytical expression for the equivalence classes of a disordered ring model.
Next, using a model of ion transport in membranes, I illustrate the symmetries of the response coefficients when multiple currents are present.
I hope these examples will help to build physical intuition and kickstart the broader application of this framework, e.g. to the design and control of nonequilibrium systems.

\section{Hidden Structure: Transport on a Disordered Ring example}

Let's consider a $N$-state system characterized by the allowed transitions $i \rightarrow  i \pm 1$ with periodic boundary conditions:
\bea
P&=&
\begin{pmatrix}
0 &  p_1 &  & &  & 1-p_1\\
1-p_2 & 0 & p_2 &  &  & \\
 &  1-p_3 & 0 & p_3 &  & \\
 &   & \ddots & 0 & \ddots & \\
 &   &  & 1-p_{N-1} & 0 & p_{N-1}\\
p_N &   &  & & 1-p_N & 0
\end{pmatrix} \, .
\label{DR}
\eea

The space of possible dynamics constitues a $N$-dimensional cube $\Sigma = \{0 < p_i \leq 1\}$ with $i =1,\cdots,N$.
Within this space, the set of equilibrium dynamics forms an $(N-1)$-dimensional hypersurface defined by
\bea
\prod_{i=1}^N \frac{p_i}{(1-p_i)} = 1 \, . 
\label{DR.EQ}
\eea
Outside of this hypersurface, the affinity
\bea
A = \ln \prod_{i=1}^N \frac{p_i}{(1-p_i)} 
\eea
differs from zero and the system is out of equilibrium. 
We see from this expression that the nonequilibrium response is not uniquely defined: An infinite number of paths exist in $\Sigma$ with the same affinities (and starting from the same equilibrium state) but with different currents \cite{A23c}.

To address this ambiguity, we can partition the space $\Sigma$ into dynamical equivalent classes as follows \cite{A12b}. 
Two dynamics $P$ and $Q$ belong to the same class if there exists a scalar $\rho \geq 0$ such that:	
\bea
P \sim Q \quad {\rm if} \quad P\circ P^{T} = \rho \, \, Q \circ Q^{T} \, ,
\label{EC}
\eea
where $\circ$ denotes the Hadamard product. 

We can use relation (\ref{EC}) to generate the equivalence classes of the disordered ring (\ref{DR}).
Consider an equilibrium dynamics $\bar{Q}$ with transition probabilities $\bar{q}_i$ satisfying (\ref{DR.EQ}).
Any dynamics $P$ in $[\bar{Q}]$ then satisfies the system of $N$ polynomial equations
\bea
p_i (1-p_{i+1}) = \rho \, \bar{q}_i (1-\bar{q}_{i+1}) 
\label{DR.pol} 
\eea
for some value of the parameter $\rho$.

This polynomial system can be solved using Gr{\"o}bner bases.
For $N=3$, the solution of (\ref{DR.pol}) takes a simple form. 
Introducing the quantities
\bea
a_i (\rho) \equiv \rho \, \bar{q}_i (1-\bar{q}_{i+1}) -1 \, ,
\eea 
the equivalence class $[\bar{Q}]$ is given by the set of dynamics
\bea
p^{\pm}_i (\rho) = \frac{1}{2a_{i+1}} (\pm \sqrt{\Delta} + a_{i-1} -a_{i}+a_{i+1})
\label{DR.EC}
\eea
with
\bea
\Delta = (a_{1}^2 + a_{2}^2 + a_{3}^2) - 2 (a_{1}a_{2}+a_{2}a_{3}+a_{1}a_{3}) - 4\, a_{1}a_{2}a_{3} \, ,
\label{DR.par}
\eea
and where $\rho$ varies between $[0,1]$ and the coefficients $a_i$ satisfy periodic boundary conditions, i.e. $a_{N+1} = a_1$ and $a_0 = a_N$. 
In this parametrization, equilibrium occurs when $\rho = 1$ or equivalently $\Delta = 0$ \cite{FN03}.
For $\rho <1$ ($\Delta > 0$), both solutions $p^{\pm}_i$ belong to $[\bar{P}]$. 
They are time-reversed of each other, with opposite affinities and currents.

\begin{figure}[h]
\centerline{\includegraphics[width=9cm]{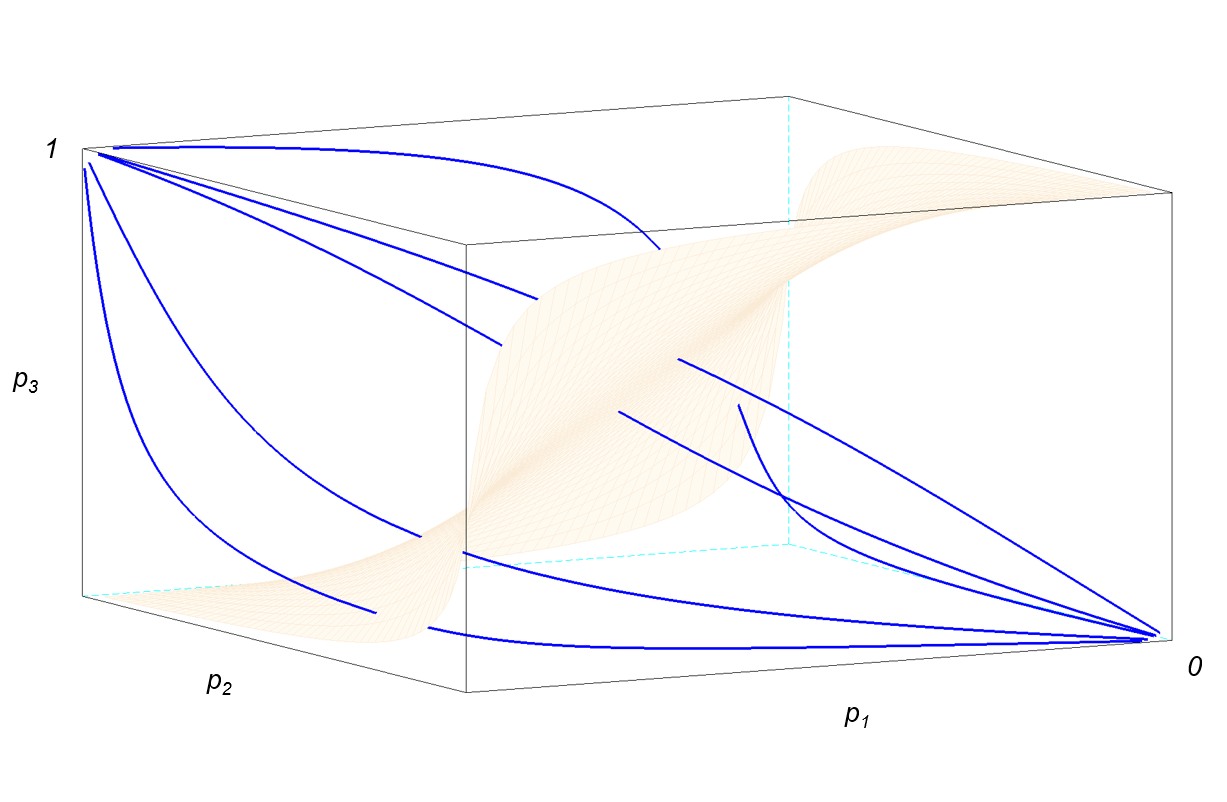}}
\caption{{\bf The space $\Sigma$ for $N=3$ and its decomposition into equivalence classes}. 
Each curve corresponds to a different equivalence class $[\bar{Q}$]. 
Each class crosses the equilibrium hypersurface once and spans the full range of affinities.}
\label{fig1}
\end{figure}

Dynamical equivalence classes thus have a clear physical interpretation.
Each equivalence class determines a curve in $\Sigma$ (more generally a manifold) that crosses the equilibrium hypersurface only once and whose affinities span the entire range $[-\infty,\infty]$ as the curves tends to $p_i = 0$ or $p_i = 1$ for all $i$ (Figure \ref{fig1}). 

As a result, within an equivalence class the nonequilibrium behavior is well-defined: there is a unique way to vary the parameters while driving the system out of equilibrium.
Moreover, for each dynamics in $[\bar{Q}]$ its time-reversed one also belongs to $[\bar{Q}]$ with opposite affinities and currents. 
These properties are valid for the equivalence classes of an arbitrary system \cite{A22, A23c}.\\

Now that we understand the structure of equivalence classes and their physical interpretation, let's use them to study the nonequilibrium behavior of the disordered ring system.

\begin{figure}[h]
\centerline{\includegraphics[width=14cm]{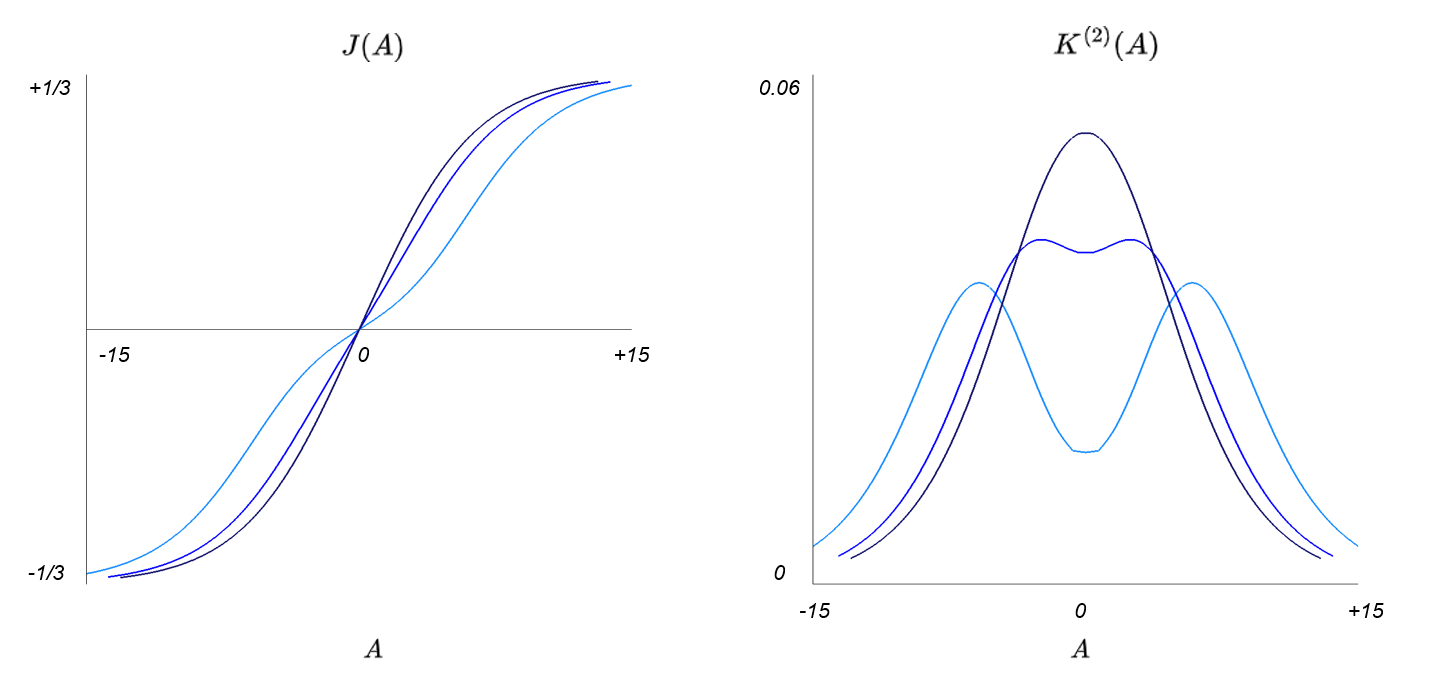}}
\caption{{\bf Current $J(A)$ and diffusion coefficient $K^{(2)}(A)$ for various equivalent classes}. 
(a) The current reaches the maximal possible value $J = \pm 1/3$ as $A$ tends to $\pm \infty$. 
(b) The diffusion coefficients are derived from the currents via Eq. (\ref{DR.fluct}). 
They vanish as the affinity goes to $\pm \infty$ in all equivalence classes. 
In contrast, varying affinities outside equivalence classes leads to different behaviors and artifacts related to the specific choice of the nonequilibrium path \cite{A23c} (not shown).} 
\label{fig2}
\end{figure}

The current $J(A)$ is illustrated in Fig. \ref{fig2}a for different equivalence classes (exact expressions for the currents can be obtained using the techniques from Ref. \cite{S76, H05, JQQ04}). 
It is antisymmetric, $J(-A) = -J(A)$, and reaches the maximal possible value $J = \pm 1/3$ as $A$ tends to $\pm \infty$.

The higher-order cumulants $K^{(n)} (A)$ can then be directly obtained from the average current as
\bea
\frac{1}{2^{n-1}} K^{(n)}(A) = \frac{{\rm d}J^n (A) }{{\rm d} A } \, .
\label{DR.fluct}
\eea
This relation holds outside the linear regime and at all orders $n$. 
In addition, knowledge of the current $J$ fully determines all the cumulants $K^{(n)}$.
In contrast, the traditional fluctuation relations determine only half of the degrees of freedom and lead to more complex relationships across cumulants \cite{A22, BG18}.

The diffusion coefficients are symmetric, $K^{(2)}(-A) = K^{(2)} (A)$, and vanish as the affinity goes to $\pm \infty$ in all equivalence classes (Figure \ref{fig2}b).
In contrast, artifacts can occur when driving a system out of equilibirum through arbitrary paths in $\Sigma$, such as saturating currents before they reach their maximal possible value, loss of their anti-symmetry, or the appearances of additional local structures \cite{A23c}.\\

The solution (\ref{DR.par}) of the disordered ring provides insights into the nonequilibrium dynamics of systems with one independent current. 
For systems for more than one currents, the equivalence classes become higher-dimensional manifolds where additional symmetries emerge \cite{A22}. 
We illustrate these additional symmetries in the next section.

\section{Hidden symmetries: Ion Transport example}

\begin{figure}[h]
\centerline{\includegraphics[width=12cm]{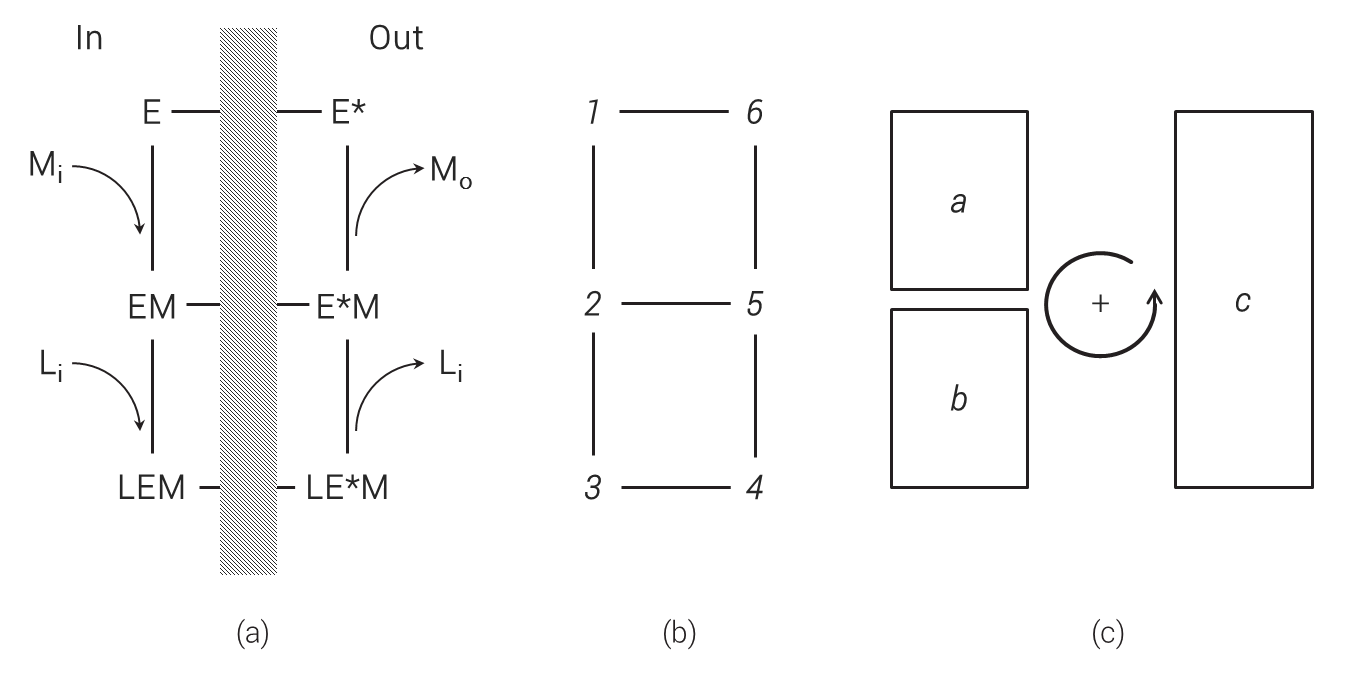}}
\caption{{\bf Ion transport model with two independent currents}. (a) Mechanism for the transport of ions M and L across the membrane. In the normal mode of operation, molecule M has a larger concentration inside than outside, $[{\rm M_i}] > [{\rm M_o}]$, while the opposite holds for molecule L, $[{\rm L_o}] > [{\rm L_i}]$. 
The complex E acts as a free energy transducer and utilizes the M concentration gradient to drive molecules of L from inside to outside against its concentration gradient. 
For example, in the case of the Na/K-ATPase complex M and L would correspond to ${\rm K}^+$ and ${\rm Na}^+$, and transport would be coupled to ATP consumption. (b) Kinetic diagram. (c) Cycle decomposition. The cycles $a$ and $b$ are chosen as the two independent cycles. The positive orientation is chosen counterclockwise (adapted from Hill \cite{H05}).}
\label{fig4}
\end{figure}

Let's now turn to a model of ion transport as an archetype of systems with two independent currents. 
Following Hill \cite{H05}, consider a cell surrounded by a membrane that separates the cell's interior (In) from its environment (Out). 
A complex E, which can exist in two conformations E and E*, has binding sites for ions L and M.
These sites are accessible to inside molecules in configuration E only, and to outside molecules in configuration E* only. 
L can be bound only when M is already bound on its site (Figure \ref{fig4}a).

The system has $N=6$ states connected by the kinetic diagram depicted in Figure \ref{fig4}b. 
The parameter space $\Sigma$ is $8$-dimensional while the set of equilibrium dynamics forms an $6$-dimensional manifold in $\Sigma$. 

\begin{figure}[h]
\centerline{\includegraphics[width=12cm]{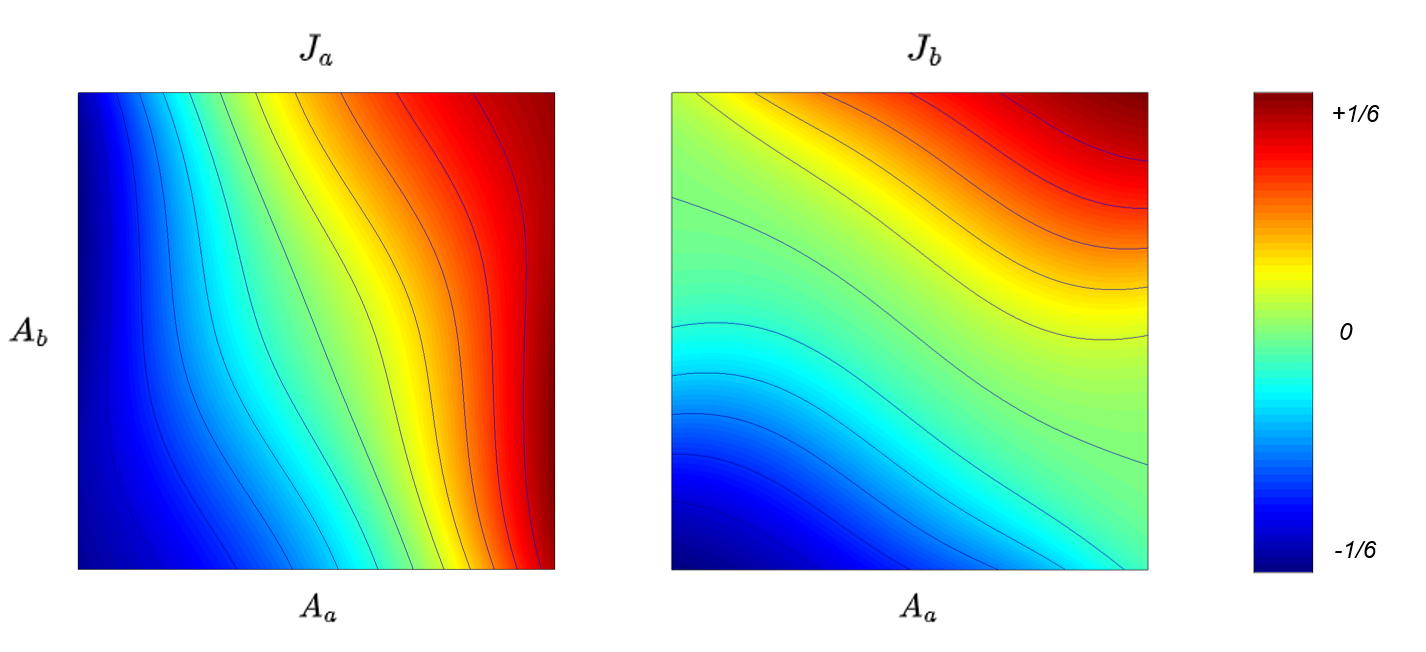}}
\caption{{\bf Coupled currents of the ion transport model.} Currents $J_a$ and $J_b$ and their isolines (solid curves) as a function of the affinities $A_a$ and $A_b$ in a given equivalence class. The equivalence class is defined by the equilibrium dynamics $p_{12} = 0.54, p_{16} = 0.46,p_{21} = 0.9, p_{23} = 0.04, p_{25} = 0.06, p_{32} = 0.06, p_{34} = 0.94, p_{43} =0.65, p_{45} = 0.35, p_{52} = 0.06, p_{54} = 0.38, p_{56} = 0.56, p_{61} = 0.58,$ and $p_{16} = 0.42$. The affinities $\pmb{A}$ take values in the range $[-10,+10]$.}
\label{fig5}
\end{figure}

Each equivalence class corresponds to a $2$-dimensional manifold \cite{FN02, FN05} related to the two independent affinties or currents (here measured along cycles $a$ and $b$, see Figure \ref{fig4}c). 
As expected, these currents are coupled, i.e. a change in $A_a$ will drive the current $J_b$ and vice versa (Figure \ref{fig5}). 
They display a nonlinear behavior and, beyond the anti-symmetry $\pmb{J}(-\pmb{A}) = - \pmb{J}(\pmb{A})$ characteristic of equivalence classes, no other symmetry is apparent. 

\begin{figure}[h]
\centerline{\includegraphics[width=15cm]{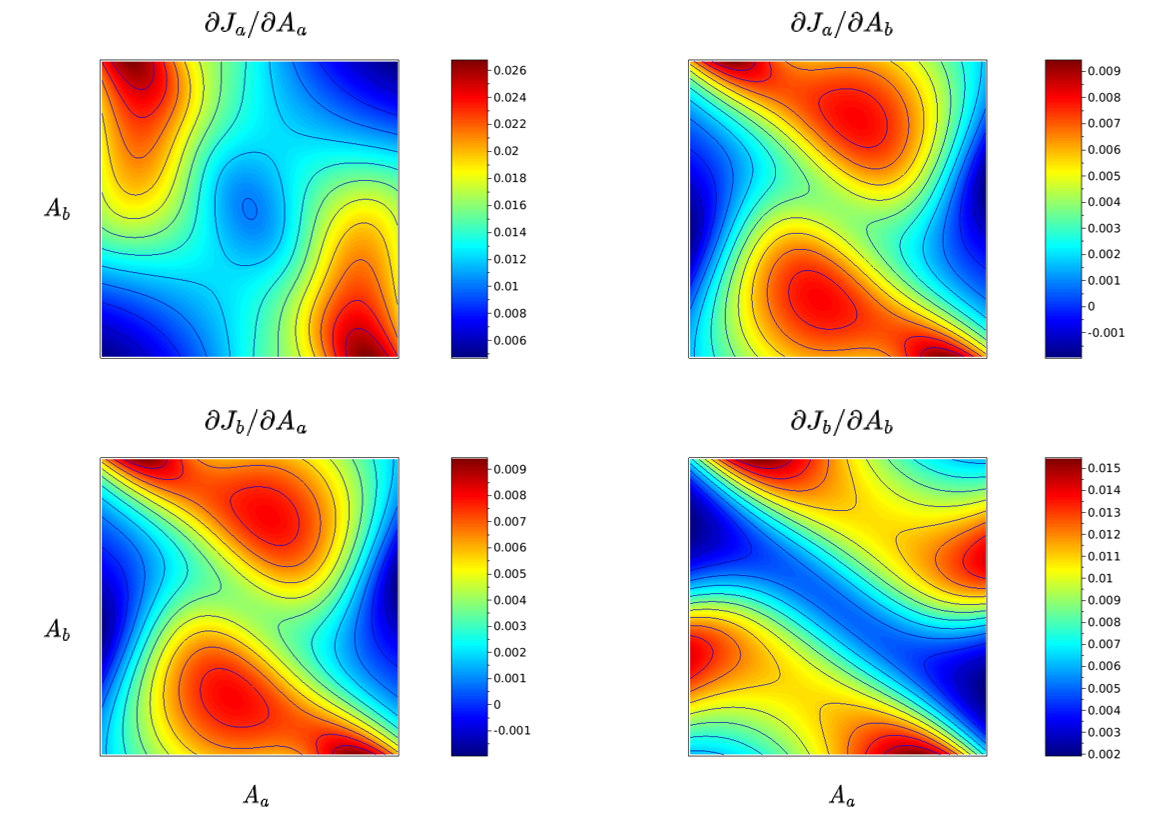}}
\caption{{\bf The currents response is symmetric both in the linear and nonlinear regimes.} The response coefficients $\partial \pmb{J}/\partial \pmb{A}$ display a complex structure in the nonlinear regime. Even so, the cross-response coefficients are identical both near and far from equilibrium. Parameters take the same values as in Fig. (\ref{fig5}).}
\label{fig6}
\end{figure}

Despite the nonlinear behavior of the currents, their response displays a 'hidden symmetry' both near and far from equilibrium \cite{A23c}:
\bea
\frac{\partial J_a}{\partial A_b} (\pmb{A}) = \frac{\partial J_b}{\partial A_a} (\pmb{A})
\label{IT.sym}
\eea
{\it for all values of the affinities $\pmb{A}$} in a given equivalence class (Figure \ref{fig6}). 
The symmetry (\ref{IT.sym}) is confirmed for all other equivalence classes (a Scilab code to generate the response coefficients for an arbitrary equivalence class is available upon request).
Systems with more than two independent currents display the same symmetries. 
Higher-order fluctuations can be obtained by further derivation of the currents and their responses are also symmetric \cite{A23c}. 

\section{Outlook}

Properties such as (\ref{DR.fluct}) and (\ref{IT.sym}) confirm the importance of dynamical equivalence classes as a central concept in nonequilibrium thermodynamics. 
Where do these remarkable properties come from?

In the traditional way of looking at nonequilibrium systems, affinities are varied along an arbitrary path in the parameter space (typically by keeping some parameters constant while varying the others). 
This makes each dynamics along the path unrelated to each other and with possibly very different kinetic properties. 
In contrast, within an equivalence class, all systems share similar kinetic properties through relation (\ref{EC}). 
Their current generating functions all have the same shapes, modulo a translation \cite{A12b, A12c}. 
All the fluctuations and response coefficients are thus directly related to each other, and equilibrium and nonequilibrium fluctuations are essentially identical.

Beyond their theoretical interests, dynamical equivalence classes bring intriguing perspectives for the design and control of nonequilibrium systems.
For example, designing mesoscopic devices around the symmetry (\ref{IT.sym}) could make them easier to control and/ or robust to perturbations.

More generally, suppose you operate a mesoscopic device designed to achieve a certain output (i.e., currents). 
The device's parameters presumably reflect a trade-off around the desired transport properties and the admissible fluctuactions and dissipation.
Imagine now that you need to achieve a different output while minimizing any other disruption to the device. 
How would you ensure the devive continues to operate as smoothly as possible?

One way to go about it would be to change the device's outputs while also staying as close as possible to the starting dynamics to minimize unrelated deviations or disruptions. 
This is precisely what dynamical equivalence classes accomplish! 
If the device has starting dynamics $G$, its equivalence class is obtained as \cite{A12a}
\bea
\min_P \left[  D(P|G) + \sum_a \lambda_a J_a [P] \right] \, .
\label{minim}
\eea
Here the $\lambda_a$s act as control parameters or Lagrange multipliers for the constrained currents. 
Meanwhile, the divergence $D(P|G) = \sum_{i,j} \pi_i P_{ij} \ln (P_{ij}/G_{ij})$ measures the statistical distance between $P$ and $G$ (in information theoric terms, $D$ is the additional cost in nats required to encode $P$ using measure $G$). 

When solving the minimization problem, the Lagrange multipliers relate to the affinities as $\lambda_a = (A_a -\bar{A}_a)/2$, where $\bar{A}_a$ are the affinities of $G$.
In other words, the minimization happens across the space of dynamics $P$ with given affinities $A_a = 2\lambda_a + \bar{A}_a$. 

Putting everything together, Eq. (\ref{minim}) reveals that equivalence classes select the dynamics minimizing the statistical distance to $G$ given fixed currents (or, equivalently, fixed affinities). 
In other words, using equivalence classes we can control a device's output while minimizing its distortions with respect to a target dynamics $G$. I believe this has practical applications and I hope these observations will stimulate further exploration of these concepts.

\vskip 1 cm

{\bf Disclaimer.} This paper is not intended for journal publication.



\begin{thebibliography}{99}

\bibitem{FN01} For example, a Markov chain with $N$ states, $E$ edges, and $S$ self-transitions will have $E-N+1$ independent currents or affinities but $2E+S-N > E-N+1$ independent parameters.

\bibitem{H05} T. L. Hill, {\it Free Energy Transduction and Biochemical Cycle Kinetics} (Dover, 2005).

\bibitem{A12b} D. Andrieux, arXiv:1208.5699 (2012).

\bibitem{A12c} D. Andrieux, arXiv:1212.1807 (2012).

\bibitem{A22} D. Andrieux, arXiv:2205.10784 (2022).

\bibitem{A23c} D. Andrieux, arXiv:2306.01445 (2023).

\bibitem{FN03} The parameter $\rho$ takes values in $[0,1]$ since I choose an equilibrium dynamics $\bar{Q}$ as the reference dynamics. Other choices would lead to a different range of values $\rho$ for which the polynomial system has real solutions \cite{A12b}. Incidently, $\rho$ is directly related to the current generating function $q = - (1/2) \log \rho$. 

\bibitem{S76} J. Schnakenberg, Rev. Mod. Phys {\bf 48}, 571 (1976).

\bibitem{JQQ04} D.-Q. Jiang, M. Qian, and M.-P. Qian, {\it Mathematical Theory of Nonequilibrium Steady States} (Springer, Berlin, 2004).

\bibitem{FN04} Using the results from \cite{S11, A12a}, the equilibrium dynamics $\bar{Q}$ of an equivalence class $[P]$ minimizes the Kullback-Leibler divergence between $\bar{Q}$ and $P$, i.e. $\bar{Q} = \min_G D_{KL}(G||P)$ where $G$ belongs to the set of compatible equilibrium dynamics.

\bibitem{S11} Shi-Feng Shieh, Entropy {\bf 13}, 2036 (2011).

\bibitem{A12a} D. Andrieux, arXiv:1201.1232 (2012).

\bibitem{FN02} For a Markov chain with $N$ states, $E$ edges, and $S$ self-transitions, the number of independent parameters is $2E+S-N$, the equilibrium hyper-surface has dimension $E+S-1$, and an equivalence class forms a $(E-N+1)$-dimensional space. 

\bibitem{FN05} The kinetic diagram (\ref{fig5}) has 14 transition probabilities $p_{ij}$. However, since they must satisfy $\sum_j p_{ij} = 1$ only 8 of them are independent. A neat decomposition of transition matrices into independent parameters can be obtained using the rotational representation of Markov chains \cite{A82, K06}.

\bibitem{A82} S. Alpern, The Annals of Probability {\bf 11}, 789 (1983).

\bibitem{K06} S. Kalpazidou, {\it Cycle Representations of Markov Processes} (Springer, 2006).

\bibitem{BG18} M. Barbier and P. Gaspard, J. Phys. A: Math. Theor. {\bf 51} 355001 (2018).

\end{thebibliography}
\end{document}